\documentclass[preprint,preprintnumbers,amsmath,amssymb]{revtex4}
\usepackage{amsfonts}    % list packages between braces
\usepackage{amssymb}
\usepackage{latexsym}
\usepackage{graphicx}
\usepackage[usenames,dvipsnames]{color}

\newcommand{\al}{\alpha}

\newcommand{\la}{\lambda}

\newcommand{\De}{\Delta}

\newcommand{\rar}{\rightarrow}

\begin{document}

\title{On $1/Z$ expansion, the critical charge for a two-electron system and the Kato theorem}

\author{A.V.~Turbiner}
\email{turbiner@nucleares.unam.mx}

\author{J.C.~Lopez Vieyra}
\email{vieyra@nucleares.unam.mx}
\affiliation{Instituto de Ciencias Nucleares, Universidad Nacional
Aut\'onoma de M\'exico, Apartado Postal 70-543, 04510 M\'exico,
D.F., Mexico}

\vskip 2cm

\begin{abstract}
\vskip 1cm

The $1/Z$-expansion for the ground state energy of the Coulomb system of an infinitely massive center of charge Z and
two electrons (two electron ionic sequence) is studied.
A critical analysis of the $1/Z$ coefficients presented in Baker et al, {\em Phys. Rev. \bf A41}, 1247 (1990) is performed and its numerical deficiency is indicated, leading, in particular, to unreliable decimal digits beyond digits 11-12 of the first coefficients. We made a consistency check of the $1/Z$-expansion with accurate energies for $Z = 1 - 10$\,: the weighted partial sums of the $1/Z$-expansion with Baker et al. coefficients, reproduce systematically the ground state energies of two-electron ions with $Z \geq 2$ up to 12 decimal digits and for $Z=1$ up to 10 decimal digits.
This rules out the presence of non-analytic terms at $Z=\infty$ contributing into the first 10-12 decimal digits in the ground state energy; it agrees with the Kato theorem about convergence of the $1/Z$-expansion within that accuracy.  The ground state energy of two-electron ions $Z=11\ (Na^{9+})$ and $Z=12\ (Mg^{10+})$ is calculated with 12 decimal digits.

\vskip 2cm

\centerline{May 31, 2015}

\end{abstract}


\maketitle

A two electron system with an infinitely-massive charge center $Z$ is described by the Hamiltonian
\begin{equation}
\label{H}
    {\cal H}\ =\ -\frac{1}{2} (\De_1 + \De_2) \ -\ \frac{Z}{r_{1}}\ -\
    \frac{Z}{r_{2}}  \ +\ \frac{1}{r_{12}}\ .
\end{equation}
A change of variables in (\ref{H}), $\vec{r} \rar \vec{r}/Z$,\ leads to a new form of the Hamiltonian
\begin{equation}
\label{H_t}
    {\cal H}_t\ =\ -\frac{1}{2} (\De_1 + \De_2)\ -\ \frac{1}{r_{1}}\ -\
    \frac{1}{r_{2}}  \ +\  \frac{\la}{r_{12}}\ , \quad \la = \frac{1}{Z}\ ,
\end{equation}
where the new energy $\tilde E(\la)=\frac{E(Z)}{Z^2}$. We will use atomic units throughout omitting dimensions.

One of the important tools to study the spectra of (\ref{H}), proposed in the early days of quantum mechanics, is
to develop the perturbation theory in (\ref{H_t}) in powers of $\frac{1}{Z}$ constructing the expansion of $\tilde E$\ \cite{Hylleraas:1930},
\begin{equation}
\label{E_t}
    \tilde E\ =\ \sum_{n=0}^{\infty} e_n \la^n\ .
\end{equation}
This is the celebrated $1/Z$-expansion.
The main reason for the interest in the $1/Z$-expansion comes from the point of view of theory: it is among very few convergent(!) expansions in quantum physics -- its convergence was proved rigorously by Kato \cite{Kato}.
It is considered as a challenge to find its radius of convergence $\la_*$ and the asymptotic behavior of
coefficients $e_n$at $n \rar \infty$. In this paper we will consider the expansion (\ref{E_t}) for the ground state only.

The first two coefficients in (\ref{E_t}) are found analytically, $e_0 = -1, e_1 = \frac{5}{8}$\, .
The first attempt to calculate the next three coefficients was carried out by Hylleraas \cite{Hylleraas:1930}.
Then many workers dedicated considerable efforts to find as many of these coefficients as possible with the highest
possible accuracy (see \cite{Silverman:1981} and \cite{Baker:1990} where extensive discussion with extended bibliography and historical account is presented) hoping to deduce the radius of convergence.
A culmination of this story happened at 1990 when Baker et al. \cite{Baker:1990} computed as many as
$\sim 401$ coefficients of $1/Z$-expansion (\ref{E_t}) essentially overpassing all previous calculations in both
accuracy and the number of coefficients.

Many years ago F.~Stillinger \cite{Stillinger:1966} presented arguments that $\la_* > 1/Z_{cr}$\ - the inverse critical charge for which the system $(Z, 2e)$ at $Z < Z_{cr}$ gets unbound, thus, $\tilde E(Z_{cr})=-\frac{1}{2}$ and
the ionization energy is equal to zero. Later this statement was challenged by W.~Reinhardt \cite{Reinhardt:1977} who conjectured that based on a possible dilatation analyticity of the problem the radius of convergence coincides
with the inverse critical charge $\la_*=1/Z_{cr}$\ and that there exists a singularity at real $\la = \la_*$\ (see for discussion \cite{Baker:1990} and references therein).

We are not aware of any further calculations of the $1/Z$-expansion coefficients performed after the paper of Baker et al, \cite{Baker:1990} was published, for the next twenty years. In 2010 the results of \cite{Baker:1990} were challenged in \cite{Zamastil:2010}. In \cite{Zamastil:2010} it was shown that the asymptotic behavior of the coefficients $e_n$ at
$n \rar \infty$ derived from the analysis of the coefficients $e_n$, taken from $n=13$ to $n=19$, differs from one
obtained from the analysis of the coefficients $e_n$ taken from $n=25$ to $n=401$ \cite{Baker:1990}. It leads to a significant deviation at large $n$ coefficients, e.g. at $n=200$ the discrepancy in the leading significant digit
is about 50$\%$ (while $e_{200}$ is itself of the order $10^{-16}$\,, see e.g. Table II in \cite{TG:2011}).
In \cite{TG:2011} this result was considered as an indication that the computational accuracy in \cite{Baker:1990}
is exaggerated, in particular, the quadruple precision arithmetics either did not work or might be insufficient,
at least, for the calculation of the first significant digits in the highest order coefficients. From another side,
in \cite{LT:2013} it was indicated that the sum of printed $e_n$-coefficients in \cite{Baker:1990} does not correspond
to the number, which was reported, in the 13-15 decimal digits. It was also shown that making a phenomenological
analysis by fitting the coefficients $e_{2-6}$ allows us to reproduce the ground state energies for $Z=1 - 10$ found accurately in \cite{Nakashima:2007} with 14 decimal digits. Those fitted coefficients $e_{2-6}$ coincided with ones
from \cite{Baker:1990} in 12 decimal digits but usually differed in all others. In particular, it was shown that
in $e_{2}$ the first 13 decimal digits obtained in the fit coincide with ones printed in \cite{Baker:1990}, when
the 14th decimal digit obtained in the fit should be 5 instead of 4\,. All the above was considered as an indication
that $e_n$-coefficients found in \cite{Baker:1990} are doubtful beyond 12 decimal digits.

Note that in \cite{TG:2011}, a terminated Puiseux expansion (in fractional degrees) was also constructed
as a high accuracy interpolation of the ground state energy close and above the critical charge $Z_{cr}$.
It was shown that the Puiseux expansion contains (with high accuracy) the integer and half-integer degrees
(exponents) and the asymptotic behavior of the $e_n$ coefficients derived from that expansion is incompatible
with the asymptotic behavior found in \cite{Baker:1990}. Also it was demonstrated that the critical charge
$\la_*$ derived from the Puiseux expansion differs in the third decimal digit from one found variationally in \cite{Baker:1990} (and recently, in \cite{Drake:2014}).

The main goal of this paper is to check compatibility of the ground state energies at $Z=1, 2, 3, \ldots , 10$
found perturbatively using partial sums (\ref{E_t}) with the coefficients $e_n$ obtained (and printed) in
\cite{Baker:1990}, with highly accurate results for the ground state energies of two-electron ions
obtained in \cite{Nakashima:2007}. It will allow us to check constructively a validity of the Kato
theorem about finiteness of the radius of convergence of the $1/Z$-expansion.
In particular, we want to calculate the ground state energies at $Z=11,12$
perturbatively, and with a maximal accuracy improving the results long ago known in literature
\cite{Thakkar:1977}. A realization of this goal implies a critical analysis of known $1/Z$-expansion
coefficients collected and printed in \cite{Baker:1990}.
Another goal is to make a perturbative calculation of the threshold energy $E_{th}$ which corresponds
to the critical charge, {\it i.e.} for which the ionization energy vanishes. Our third goal is to try
to construct a terminated Puiseux expansion with integer and half-integer degrees assuming a presence
of a singularity at the critical charge {\it exactly} (reliably calculated in \cite{Drake:2014}).
All calculations are going to be cross-checked in two different multiple precision arithmetics:
(i) Intel ifort q-precision real*16 (quadruple precision) and (ii) Maple Digits=30 in Maple 13.

We begin our presentation from a critical analysis of the $1/Z$-expansion coefficients $e_n$
collected and printed in \cite{Baker:1990} indicating three numerically incorrect statements.

(I).\ The first observation is that we do not confirm the statement from \cite{Baker:1990} (p.1254)
(see \cite{LT:2013}):

\noindent
\textit{The sum of the $e_n$'s for $n$ running from 0 to 401 is
\begin{equation}
\label{EB1}
 -0.527 \, 751 \, 016 \, 544 \, 266
\end{equation}
which, at the time we did our calculations, was the most accurate estimate of the energy for the ground state of H$^-$}.
Our own result of summation using $e_n$ printed in \cite{Baker:1990}, Table III with ten significant figures (eventually presented with , at least, 12 decimals and sometimes with even more decimals)
\footnote{We added zeroes afterwards to have a total of 15 decimals in each coefficient. Those digits were neither printed
nor mentioned in \cite{Baker:1990}, but needed for consistency in order to get 15 decimal digits in the sum (3), in order to compare it with the one printed in (\ref{EB1})}
\[
 -0.527 \, 751 \, 016 \, 544 \, 160\ .
\]
It differs in the last three decimal digits. It gives us a reason to suspect that either the quadruple precision arithmetics used by the authors of \cite{Baker:1990} was insufficient due to of error accumulation, or the length of the trial function expansion (476 basis functions) was not enough, or both factors together, prevent going beyond 12 decimal digits (or non-printed decimals which we put equal to zero are essential, see below a discussion). Thus, we can not trust decimal digits beyond 12th in coefficients $e_n$. In particular, all $e_n$ for $n > 135$ (when the (rounded) coefficients are of the order $10^{-12}$ and less, see \cite{Baker:1990}, Table III) seem untrustable.

In conclusion, we have to note that both the above numbers for energy at $Z=1$ coincide up to 12 decimal digits with the accurate result obtained in \cite{Nakashima:2007}
\begin{equation}
\label{EN1}
  -0.527 \, 751 \, 016 \, 544 \, 377\ \mbox(rounded)
\end{equation}
but differ from it in 13th and subsequent decimal digits. It is the explicit indication that $e_n$ (all or some) calculated in \cite{Baker:1990} beyond 12 decimal digits are not trustful. Furthermore, making a comparison of the numbers (\ref{EB1}) and (\ref{EN1}) one can draw a conclusion that the results obtained in \cite{Baker:1990} remain doubtful even if we assume that more significant figures in $e_n$ were calculated but for whatever reason were not printed in \cite{Baker:1990} but were used to obtain the sum (\ref{EB1}). In general, the result (\ref{EB1}) contradicts the Kato's theorem on the convergence of the $1/Z$ expansion beyond twelve decimal digits.

(II).\  The second observation is that we do not confirm another statement
from \cite{Baker:1990} (p.1254):

\noindent
\textit{ For $Z=2$ the corresponding weighted sum of coefficients ... yields an estimate of
\begin{equation}
\label{EB2}
-2.903\,724\, 377\, 034\, 116\, 7
\end{equation}
for the ground state energy of helium \ldots}. Our own result {using $e_n$ printed in \cite{Baker:1990} with ten significant figures, in general, but with, at least, 12 decimals (by adding zeroes afterwards, see footnote [16])}
\[
 -2.903\, 724\, 377\, 034\, 051\, 9 \quad \mbox(non-rounded)
\]
differs in the last four decimal digits. We have to note that both the above numbers coincide up to 12 decimal digits (before rounding) with accurate result given in \cite{Nakashima:2007} for $Z=2$
\begin{equation}
\label{EN2}
      -2.903 \, 724 \, 377 \, 034 \, 119 \, 6 \ \mbox(rounded)
\end{equation}
Even if we assume that more significant figures in $e_n$ were calculated but not printed in \cite{Baker:1990} the result (\ref{EB2}) differs from (\ref{EN2}) in the 15-16th decimal digits. This  is one more indication (cf (I)) that $e_n$ calculated in \cite{Baker:1990} beyond the 14th decimal digit are not trustful. In general, the result (\ref{EB2}) contradicts the Kato's theorem on the convergence of the $1/Z$ expansion beyond fourteenth decimal digit.

Following the above observations (I) and (II) the number of trustable {\it significant} digits in $e_n$ from \cite{Baker:1990} reduces gradually with increase of $n$ and becomes for $e_{40, 50}$ (where the first seven decimal digits are zeroes) equal to 5, out of 12 {\it decimal} digits. It leads to questioning the statement from \cite{Baker:1990} (p.1254):

(III).\ {\it The results ... suggest that even our higher order
$e_n$'s are accurate to a few parts in $10^{-5}$ and our high-order $r_n$'s {\rm (the ratio of subsequent coefficients)} to a few parts in $10^{-6}$.}

It implies, in particular, that even the first significant digit in the coefficients $e_n$ at $n > 135$ is not trustable.
These coefficients form a great portion of data used for interpolation and eventually for the extraction of asymptotic
behavior of coefficients $e_n$\,. Hence, the obtained asymptotic behavior in \cite{Baker:1990} is not fully trustable.

Keeping in mind the observations (I)-(III) we take the coefficients $e_n$ printed in \cite{Baker:1990} (see Table III therein)
and calculate the partial weighted sum of (\ref{E_t}) up to $n=401$ for $Z=1 \ldots 10$.
\begin{table}[hbt]
\begin{center}
{\small
\begin{tabular}
{l|rr}
\hline
$Z$ & $E$ (a.u.) from (\ref{E_t}) \quad &\quad $E$ (a.u.) \\
\hline
$Z^{EBMD}_{cr}$ &  -0.414 986 047  \ &\ -0.414 986 212 532 679  \\
% 196 590 814 566 747 511 383  045 02 \\
    &\ \ -0.414 978 381 \quad $(*)$ \ & \ -0.414 986 212 53\ \
\cite{Horop:2014}                                    \\
1  &  -0.527 751 016 544 (2)                  & -0.527 751 016 544 377
 \\
 %   196 590 814 566 747 511 383  045 02 \\
    & \ \ -0.527 751 016 471\ $(*)$    &
  \\
2  &  -2.903 724 377 034 (1)       \ &\ -2.903 724 377 034 119       \\
%   598 311 159 245 194 404 446 696 905 \\
3  &  -7.279 913 412 669 (3)       \ &\ -7.279 913 412 669 305      \\
%   964 919 459 221 006 611 682 572 35 \\
4  &  -13.655 566 238 423 (6)      \ &\ -13.655 566 238 423 586     \\
%   702 081 730 194 612 159 391 360 60 \\
5  & -22.030 971 580 242 (8)       \ &\ -22.030 971 580 242 781     \\
%   541 655 702 043 566 870 379 775 99 \\
6  & -32.406 246 601 898 (5)       \ &\ -32.406 246 601 898 530      \\
%   310 557 357 969 530 254 566 016 97 \\
7  & -44.781 445 148 772 (7)        \ &\ -44.781 445 148 772 704      \\
%   645 185 760 848 954 056 776 028 12 \\
8  & -59.156 595 122 757 (9)        \ &\ -59.156 595 122 757 925      \\
%   558 549 892 445 559 527 700 907 85 \\
9  &  -75.531 712 363 959 (5)       \ &\ -75.531 712 363 959 491      \\
%   104 878 015 579 533 576 560 909 77 \\
10 &  -93.906 806 515 037 (5)      \ &\ -93.906 806 515 037 549      \\
%   421 469 184 180 000 241 066 651 70 \\
11 & -114.281 883 776 072 (7)    \ &\      -114.281 879\  $(**)$       \\
12 & -136.656 948 312 646 (9)   \ &\      -136.656 944\  $(**)$       \\
 \hline
\end{tabular}
}
\end{center}
\caption{\label{E}
Perturbative Theory energies $E(Z)$ from (\ref{E_t}) with coefficients
$e_n, n=0,1,\ldots 401$ from \cite{Baker:1990}, $(*)$ with reduced $\#$ of decimal digits to 12 (see text).
Energies $E$ (rounded, right column) at the threshold, for $Z=1$
(see text), from Ref. \cite{Nakashima:2007} for $Z>1$;
the results \cite{Thakkar:1977} marked by $(**)$}
\end{table}
We obtain an agreement between the energies found perturbatively using the partial sum in (\ref{E_t}) and the ones in \cite{Nakashima:2007} for all $Z=1 \ldots 10$: they coincide up to 12, sometimes, 13 decimal digits, see Table \ref{E}.
Since the contribution of the higher order coefficients into the weighted sum (\ref{E_t}) to the ground state energy decreases dramatically with the increase of the charge $Z$, it is guaranteed that the same number 12 of correct decimal digits, at least, should be obtained for larger $Z > 10$. Based on that, the ground state energies for two-electron ions $Z=11\ (Na^{9+})$ and  $Z=12\ (Mg^{10+})$ are calculated perturbatively and presented in Table \ref{E}. They improve
the most accurate results obtained in \cite{Thakkar:1977}: they differ from them in the sixth decimal digit. However,
for $Z=1$ the agreement between the energy found perturbatively through (3) and the one from \cite{Nakashima:2007}
occurs up to the 12th decimal digit  (before rounding). We made an experiment by rounding the coefficients $e_n$ from \cite{Baker:1990} to 12 decimals. It implies, in particular, that all coefficients $e_n = 0$, for $ n > 135$. Then we calculated the weighted sums  (\ref{E_t}) for different $Z$. For $Z = 2, \ldots, 12$ the results for weighted sums remain essentially unchanged up to 12 decimals, see Table \ref{E}, sometimes making a difference in one portion of $10^{-12}$. It implies that, in fact, decimals beyond 12th in $e_n$ do not give a contribution to the first 12 decimal digits in ground state energies for $Z = 2, \ldots, 12$. However, for $Z=1$ the situation becomes different: the sum is increased in such a way that a coincidence with the energy of \cite{Nakashima:2007} occurs up to the 10th decimal digit (after rounding) only, see Table \ref{E}. This can be considered as an indication to incorrectly calculated decimal digits in \cite{Baker:1990}
in the first coefficients $e_n$ and, perhaps, for $e_n, n > n_0$ beyond the first 10th decimals(!), where $n_0$ is some integer. It is certainly the indication that $e_n$-coefficients should be checked/recalculated beyond ten decimal digits
in order to get an agreement with the energy found in \cite{Nakashima:2007} for $Z=1$.

Recently, Estienne et al. \cite{Drake:2014} (EBMD) in a remarkable, high precision variational calculation with triple basis sets containing up to 2276 terms, obtained a highly accurate value for the critical charge
\begin{equation}
\label{zcrit}
      Z^{EBMD}_{cr}\ =\ 0.911\, 028\, 224\, 077\, 255\, 73\ ,
\end{equation}
and for the linear slope of $E (\la)$ at the critical charge, 0.2451890639.
This result (\ref{zcrit}) coincides in 5 decimals with the one from \cite{Baker:1990} and in 3 decimals with the one
from \cite{TG:2011} (after rounding). In spite of the Reinhardt conjecture claim that the $1/Z$ expansion diverges at $Z_{cr}$ we calculated the weighted sum (\ref{E_t}) for such a critical charge using the coefficients from
\cite{Baker:1990}, see Table \ref{E}. Surprisingly, this result when compared with the exact energy at threshold,
\begin{equation}
\label{th}
E_{th} = - \frac{(Z^{EBMD}_{cr})^2}{2}\ ,
\end{equation}
leads to a difference $\sim 10^{-7}$.  The remainder in (\ref{E_t}) is
\[
      \frac{e_{401}}{Z_{cr}^{401}} \ \sim \ -2 \times 10^{-9}\ .
\]
It indicates that the singularity at $Z_{cr}$, if it exists, is pretty "weak". Recently, the value of the critical charge (\ref{zcrit}) was calculated in a direct solution of the Schroedinger equation for the Hamiltonian (1) using the Lagrange-mesh method \cite{Horop:2014}. It was found that for $Z=Z^{EBMD}_{cr}$ the lowest eigenvalue coincides with threshold energy $E_{th}$ in 12 decimal digits. That corresponds to the ionization energy $\sim 10^{-12}$. Thus, twelve decimals in $Z^{EBMD}_{cr}$ have been verified in an independent calculation.

With accurate knowledge of the critical charge (\ref{zcrit}) and linear slope of $E (\la)$ at
the critical charge \cite{Drake:2014} (predicted in \cite{Simon:1977}) one can make the analysis of the behavior of the ground state energy $\tilde E (\la)$ in the vicinity of the inverse critical charge at $\la \leq {\la}_{cr}$ fitting $\tilde E (\la)$ via the terminated Puiseux expansion with integer and half-integer exponents (c.f. \cite{TG:2011}). This time it is the interpolation of the energy $E(Z)/Z^2$ at eight points including the critical charge, $Z \in [Z^{EBMD}_{cr}, 0.95, 1., 1.05, 1.1, 1.15, 1.2, 1.25]$ taking the energies for $Z > Z^{EBMD}_{cr}$ calculated in \cite{TG:2011}\ ,
\footnote{Note that the ground state energies were calculated in \cite{TG:2011} using the Korobov basis set of a size which leads to twelve correct significant figures in the energy for H${}^-$, at $Z=1$. However, it does not guarantee the same number of correct significant figures in energy for smaller $Z<1$. Furthemore, it was shown in \cite{Horop:2014} that the accuracy deteriorates dramatically: for $Z=0.95$ the accuracy drops to one portion $10^{-7}$(!) while for $Z > 1$ it remains equally accurate as for $Z=1$.  Since we do not require our fit to reproduce more than 6-7 decimals, this deficiency at $Z=0.95$ does not really hurt us.}\ .
This leads to the expansion
\begin{eqnarray}
 {\tilde E}^{(fit)}(\la)\ &=&\ -\frac{1}{2} - 0.2451890639\  {\tilde \la}
 - 0.0252309\ {\tilde \la}^{3/2} - 0.5532438\ {\tilde \la}^2\
\nonumber \\
  && +\ 0.9729112\ {\tilde \la}^{5/2} - 0.707285\ {\tilde \la}^3\ + \ldots \ ,
\label{e-fit_2}
\end{eqnarray}
(c.f. \cite{TG:2011}), where $\tilde \la = ({\la}^{EBMD}_{cr}-\la)$\ and ${\la}^{EBMD}_{cr} = 1.097 660 833 738 559 80$ \cite{Drake:2014}.
The expression (\ref{e-fit_2}) reproduces 7-6 s.d. in energies at $Z$ close to the critical charge, then
gradually deteriorates with the increase in $Z$ giving 3 s.d. at $Z=1.25$\,; for illustration see Table~\ref{table2}.
This fit is not of a good quality: the coefficients are large in front of terms of larger degrees; they do not demonstrate a tendency to converge. The coefficient in front of ${\tilde \la}^{3/2}$ is rather small compared to other coefficients. Probably, it can vanish if we would reduce the interval of interpolation from $[Z^{EBMD}_{cr}, 1.35]$ to say $[Z^{EBMD}_{cr}, 0.95]$. It seems as an important check of viability of the expansion (\ref{e-fit_2}): since we do not know the radius of convergence of the Puiseux expansion, we do not know a domain in $Z$ to choose for interpolation. In order to realize this check, the calculations of energies in this domain should be done in a method other than the Korobov basis set used in \cite{TG:2011} (see also footnote [17]), for example, in the Lagrange mesh method \cite{Horop:2014} or in triple basis set \cite{Drake:2014}. This will be done elsewhere.

%The existence of the Puiseux expansion (\ref{e-fit_2})
%leads to a conclusion that the energy $E(Z)$ (or $\tilde E (\la)$) has square-root singularity with exponent 3/2 at the %critical charge in agreement with \cite{Stillinger:1966}, \cite{Zamastil:2010}, \cite{TG:2011}. Thus, it can
%be considered as an indication the validity of the Reinhardt conjecture about a connection between the radius $\la_{\star}$ of %convergency of $1/Z$-expansion and the critical charge, $\la_{\star}=\frac{1}{Z_{cr}}$. Physics meaning of this singularity is %unclear: can it make sense as a point of level crossing.

\begin{table}
\begin{tabular}{|c|c|c|}
\hline\hline
Z          &\       $E$        \ &\  $E^{(fit)}$\  \\
\tableline
\ 1.3   \  &\  -0.609 406 309   \ &\  -0.609 9 \  \\
\ 1.25  \  &\  -0.597 488 174   \ &\  -0.597 7 \  \\
\ 1.15  \  &\  -0.571 655 437   \ &\  -0.571 68  \  \\
\ 1.00  \  &\  -0.527 751 017   \ &\  -0.527 751 009 \  \\
\ 0.95  \  &\  -0.512 049 529   \ &\  -0.512 049 511 \  \\
\tableline
\hline
\end{tabular}
\caption{
\label{table2}
 Ground state energy $E$ for two-electron ion for selected values of Z found in \cite{TG:2011}
 and with the correction for $Z=0.95$ for the sixth and next decimals, see \cite{Horop:2014};  here
 all the displayed digits are assumed to be correct, $E^{(fit)}$ from the fit (\ref{e-fit_2}).
}
\end{table}

To conclude, we state that the use of the $e_n$-coefficients found by Baker et al. in the weighted sum (3) allow us
to reproduce up to 12 decimals in the exact ground state energies for $Z=1,\ldots 10$. In a way it confirms constructively
the validity of theorem by Kato about a finite radius of convergence of $1/Z$ expansion on the level of 10 decimal digits
and rules out the existence of exponentially-small terms $\sim \exp (- a |Z|^{\al})$ with parameters $a>0, \al \geq 1$.
Since we have no doubts in validity of the Kato theorem, it indicates that the $e_n$-coefficients found by Baker et al.
have to be recalculated to check the validity of their higher digits, beyond 10 decimal figures.
Note that if the Feynman diagram technique for calculations of $e_n$ can be established which seems plausible,
one could expect to have $e_n$ in a form of superposition of Riemann $\zeta$-functions with integer arguments.

We feel a necessity to develop an alternative, analytical approach for finding the asymptotic behavior of
$e_n$-coefficients, probably, similar to one based on dispersion relations in a coupling constant for
anharmonic oscillators due to Bender and Wu \cite{BW}, or one derived directly from a path integral as it is done
in quantum field theory. A separate issue is to find the level crossings in the $\la$-complex plane and associated
square-root branch points, especially, closest (and close) to $\la_{cr}$, $\la=0$, and its respective
contributions to the $e_n$ coefficients. This might be the subject of a future work.

The low quality of the interpolation of the behavior of $E(\la)$ near $\la_{cr}$ using a finite number of terms
in the Puiseux expansion with integer and half-integer degrees at $\la=\la_{cr}$ might be either an indication
to the absence of a singularity at $\la_{cr}$ and terms of fractional degrees in the Puiseux expansion, or, at least, to a small radius of convergence of the Puiseux expansion.
This question should be studied more carefully, in particular, from a numerical point of view.
Without addressing above issues the validity of the Reinhardt conjecture can not be established.

The research is supported in part by DGAPA grants IN109512, IN108815 (Mexico). The authors are thankful to the participants of the seminars at IIMAS-UNAM (Mexico), AMO at University of Stony Brook and University of Connecticut, ITAMP (Harvard) for interest to work and useful remarks.

\end{document}